\documentclass{aa}

\def\kpc{{\rm kpc}}
\def\kms{{\rm km\,s^{-1}}}

\def\kmskpc{\,km\,s$^{-1}$\,kpc$^{-1}$}

\def\mVR{$\langle V_R \rangle$}
\def\mVphi{$\langle V_\phi \rangle$}
\def\VZb{$V_{\rm{Z,\,br}}$}

\newcommand{\mycomment}[1]{}

\usepackage{graphicx}
\usepackage{txfonts}
\begin{document}

   \title{Dark matter spiral arms in Milky Way-like halos}
   
   \author{Marcel Bernet \inst{1,2,3}
            %\and et al.
            \and Pau Ramos \inst{4}
            \and Teresa Antoja \inst{1,2,3}
            \and Victor P. Debattista \inst{5}
            \and Martin D. Weinberg \inst{6}
            \and\\ Jo\~{a}o A. S. Amarante \inst{7,8}
            \and Robert J. J. Grand \inst{9}
            \and {\'O}scar Jim\'{e}nez-Arranz \inst{10}
            \and Chervin F. P. Laporte \inst{11,2,12}
            \and \\Michael S. Petersen \inst{13}
            \and Santi Roca-F\`{a}brega \inst{10}
            \and Merc\`{e} Romero-G\'{o}mez \inst{1,2,3}
          }
    
    \institute{Departament de Física Qu\`antica i Astrof\'isica (FQA), Universitat de Barcelona (UB),  c. Mart\'i i Franqu\`es, 1, 08028 Barcelona, Spain
           \email{mbernet@fqa.ub.edu}
    \and{Institut de Ci\`encies del Cosmos (ICCUB), Universitat de Barcelona (UB), c. Mart\'i i Franqu\`es, 1, 08028 Barcelona, Spain}
    \and{Institut d'Estudis Espacials de Catalunya (IEEC), Edifici RDIT, Campus UPC, 08860 Castelldefels (Barcelona), Spain} 
    \and{National Astronomical Observatory of Japan, Mitaka-shi, Tokyo 181-8588, Japan}
    \and{Jeremiah Horrocks Institute, University of Central Lancashire, Preston, PR1 2HE, UK}
    \and{Department of Astronomy, University of Massachusetts at Amherst, 710 N. Pleasant St, Amherst, MA 01003, USA}
    \and{Department of Astronomy, School of Physics and Astronomy, Shanghai Jiao Tong University, 800 Dongchuan Road, Shanghai, 200240, China}
    \and{State Key Laboratory of Dark Matter Physics, School of Physics and Astronomy, Shanghai Jiao Tong University, Shanghai 200240, China}
    \and{Astrophysics Research Institute, Liverpool John Moores University, 146 Brownlow Hill, Liverpool L3 5RF, UK}
    \and{Lund Observatory, Division of Astrophysics, Lund University, Box 43, SE-221 00 Lund, Sweden}
    \and{LIRA, Observatoire de Paris, Universit\'e PSL, Sorbonne Universit\'e, Universit\'e Paris Cit\'e, CY Cergy Paris Universit\'e, CNRS, 92190 Meudon, France}
    \and{Kavli IPMU (WPI), UTIAS, The University of Tokyo, Kashiwa, Chiba 277-8583, Japan}
    \and{Institute for Astronomy, University of Edinburgh, Royal Observatory, Blackford Hill, Edinburgh EH9 3HJ, UK}
    }
   \date{Received YYY; accepted XXX}

% \abstract{}{}{}{}{}
% 5 {} token are mandatory
 
  \abstract
  % Context
  {The coupling between the dark matter (DM) halo and the stellar disc is a key factor in galactic evolution. While the interaction between structures like the Galactic bar and DM halos has been explored (e.g. slowing down of the bar due to dynamical friction), the effect of spiral arms on the DM halo distribution has received limited attention.}
  % Aim
  {We aim to detect and characterize the interaction between the stellar spiral arms and the DM halo.}
  % Methods
  {We analysed a suite of simulations featuring strong stellar spiral arms, ranging in complexity from test-particle models to fully cosmological hydrodynamical simulations. Using Fourier transforms, we mapped the phase and amplitude of the stellar spirals at different times and radii. We then applied the same methodology to DM particles near the stellar disc and compared trends in Fourier coefficients and phases between the two components.}
  % Results
  {We detect a clear spiral arm signal in the DM distribution, correlated with the stellar spirals, confirming the reaction of the halo. The strength of the DM spirals consistently measures around 10\% of that of the stellar spiral arms. In the $N$-body simulation, the DM spiral persistently trails the stellar spiral arm by approximately $10^\circ$. A strong spiral signal of a few kilometres per second appears in the radial, azimuthal, and vertical velocities of halo particles, distinct from the stellar kinematic signature. In a test-particle simulation with an analytical spiral potential (omitting self-gravity), we reproduce a similar density and kinematic response, showing that the test-particle halo responds in the same way as the $N$-body halo. This similarity confirms that we are observing the forced response of the halo to the stellar spiral arms potential. Finally, we also find the presence of DM spiral arms in a pure $N$-body simulation with an external perturber, and isolated and cosmological hydrodynamical simulations, indicating that the dynamical signatures of the forced response in the DM halo are independent of the dynamical origin of the stellar spiral arms.}
  % Conclusions
  {We reveal the ubiquitous presence of DM spiral arms in Milky Way-like galaxies, driven by a forced response to the stellar spiral potential.}

   \maketitle

\section{Introduction}

Dark matter (DM) is essential for galaxy formation and evolution, making up about 84\% of the matter of the Universe \citep{plank2020,desi2024}. It provides the gravitational framework for the appearance of galactic structure and helps explain discrepancies between observed rotation curves and the distribution of baryonic mass \citep{rubin1970dm,rubin1980dm}, large-scale structures \citep{efstathiou1985largescale}, and colliding systems such as the Bullet Cluster \citep{markevitch2004bulletcluster}. The DM halo and the stellar component of galaxies are dynamically coupled, meaning that their gravitational interactions influence each other. While the halo shapes the overall stellar structure of galaxies, the stellar component, in turn, induces gravitational responses within the halo, potentially leaving detectable signatures. Understanding these interactions improves our knowledge of DM, galactic dynamics, and the overall structure of the Universe.

One notable effect of this coupling between components is the gradual slowing down of the Galactic bar rotation, driven by the exchange of angular momentum between the stellar disc and the DM halo \citep[e.g.][]{weinberg1985slowbar,hernquist1992slowbar,debattista1998nbodyhalos,debattista2000nbodyhalos,athanassoula2003bar,chiba2022dmbar}. 
This process is a result of the resonant nature of the system. Indeed, \citet{tremaine1984slowbar} found that the exchange of angular momentum in the halo is driven by particles that are nearly in resonance.
The other classical example is the orbital coupling of satellite galaxies, governed by dynamical friction \citep{chandrasekhar1943,banik2022dynamicalfriction}, whereby satellites lose orbital energy as they move through the DM halo, eventually spiralling inward \citep[e.g.,][]{tremaine1975decay,weinberg1986decay}.

In recent years several studies have focussed on the impact of the interaction between the stellar and DM components on the precise structure of the halos. One of the most studied cases is the disturbance of the Milky Way (MW) halo caused by the infall of the LMC \citep[e.g.][]{weinberg1998transient,gomez2015lmcwarp,laporte2018lmc,garavitocamargo2019wake,cunningham2020wake,petersen2021reflex,amarante2024collective}.
In addition, early theoretical works \citep{weinberg1985slowbar,hernquist1989quadrupole} predicted that stellar bars generate a quadrupole wake in the DM halo. Later $N$-body simulations confirmed the emergence of a DM overdensity in response to the bar potential \citep[e.g.][]{weinberg2002dmbar}, though these studies primarily characterized the halo response as classical wakes.
\citet{petersen2016shadowbar} demonstrated that DM particles can become trapped in the bar potential, along with stars, forming a persistent `shadow bar' that contributes $\sim\,10\%$ of the total bar mass \citep[see also][]{collier2019dmbar,collier2021dmbar,forsst2024dmbar,marostica2024dmbar,ash2024dmbarTNG}.
Another example of interactions between components is the trapping of halo stars into bar-related resonances, proposed in \citet{dillamore2023barhalo1,dillamore2024barhalo2} as an explanation for the `chevrons' observed in \emph{Gaia} data \citep{belokurov2023chevrons}.

The stellar spiral arms and the DM halo interact similarly.
In general, the particles in the halo must respond to any perturbation in the potential. Therefore, a spiral perturbation in the potential will create a response in the halo. 
However, the nature and consequences of this response are surprisingly under-examined in the literature. \citet{mark1976} predicted a significant amplification of the stellar spiral density waves as a result of the loss of angular momentum to the halo.
Later, \citet{fuchs2004sheet} employed the shearing sheet model \citep{goldreich1965swingamplification,julian1966swingamplification} to investigate spiral amplification in the presence of a live DM halo. He predicted that the maximum growth factor would increase significantly due to the coupling with the halo and pointed out that the transfer of momentum is exclusively mediated by halo particles whose orbits are in resonance with the spiral waves. This response can be described as a broader dynamical process: a `forced response', whereby periodic perturbations (e.g. a bar, a satellite, and the spiral arms) systematically distort orbits, with the strongest effects occurring near resonances.
In the same set-up, he also studied the response of the halo to a density wave in the disc, and reported the development of a wake. Subsequent studies found that halos with anisotropic velocity distribution increase the maximum growth factor of the spirals considerably \citep{fuchs2005anisotropy}.
More recently, \citet{sellwood2021live} tested these predictions using a complete $N$-body simulation and reported that a live DM halo had no significant effect on the growth rate of the spiral modes, although he showed hints of an increase in the final stellar spiral amplitude of about $20\%$.

In low-surface-brightness galaxies (LSBs), the disc-halo coupling plays a key role in the evolution of spiral arms, due to their low stellar densities and DM-dominated dynamics. Theoretical studies have predicted that halos with a small local vertical scale height near the disc or halos with a significant angular momentum can strongly amplify disc density waves through resonant interactions. \citet{chiueh2000lowmass} proposed that such coupling mechanisms could not only excite spiral-arm formation but also influence their morphology, with tighter pitch angles associated with smaller halo scale heights. Recent simulations by \citet{narayanan2024lowmass} confirm the importance of the geometry of the halo in driving spiral patterns in LSBs, showing that oblate halos can sustain long-lived global spirals.

While earlier studies have concentrated on the feedback effects of coupling on the stellar component, our work shifts focus to the morphology of the DM halo itself. In this work, we find clear signs of a spiral-arm-shaped structure in the DM halo in several MW-like simulations with cold DM.
We start by characterizing the overdensities created in the DM halo by the stellar spiral arms in an isolated pure $N$-body simulation.
We reproduce the response of the DM halo to the gravitational perturbation induced by the stellar spiral arms using a simple test particle simulation. This demonstrates that the relation between the stellar spiral arms and the DM substructure observed in the pure $N$-body simulation is dominated by a forced response to the modification of the potential. Finally, we observe similar interaction patterns in pure $N$-body simulations with a massive perturber, isolated hydrodynamical simulations, and cosmological simulations, proving the ubiquitous presence of DM spiral arms in the state-of-the-art simulations of MW-like systems.

In this work, we show that in any MW-like system with cold DM, fairly strong spiral arms will have an increase of about $10\%$ in their total mass, due to the forced response of the DM halo. The existence of DM spirals implies that the local DM density varies non-axisymmetrically across the galactic disc and, in particular, close to the Sun. Thus, the torques in a MW-like spiral galaxy will have contributions from both the stellar and the DM components. In the future, the kinematic and density maps of the DM particles provided in our work might be used as test for the halo distribution of our Galaxy.

In Sect. \ref{sect:simulations_methods}, we briefly describe the simulations and methods used throughout the paper. In Sect. \ref{sect:results}, we present our findings for the isolated pure $N$-body model. These results are then compared with a test particle model in Sect. \ref{sect:results_tp}, where we also perform some tests on the characteristics of the response. In Sect. \ref{sect:ubiquity}, we test the presence of DM spiral arms in a set of different simulations.
These results are discussed in Sect. \ref{sect:discussion}. Finally, in Sect. \ref{sect:conclusions} we provide a summary of our results and list our conclusions.

\begin{figure*}
    \centering
    \includegraphics[width=1.\textwidth]{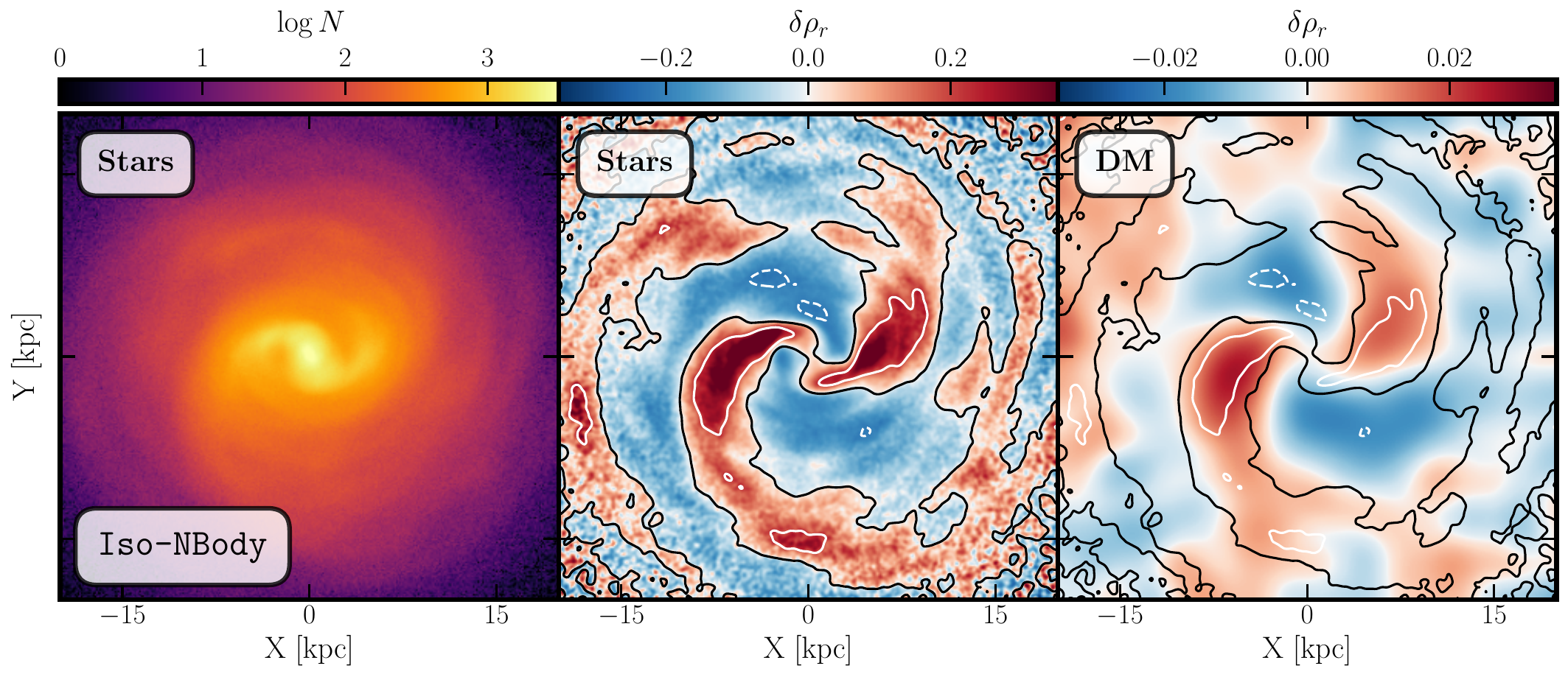}
    \caption{Spiral arms in the \texttt{Iso-NBody} simulation, at $t=7$\,Gyr. \textit{Left}: Surface density plot of the stellar component. \textit{Middle}: Radially normalized density ($\delta\rho_r$) of the stellar component. The contours represent the regions where $\delta\rho_r=0$ (in black), and $\delta\rho_r=\pm0.2$ (in white). \textit{Right}: $\delta\rho_r$ of the DM particles around the stellar disc ($|Z| < 4$\,kpc). The contours are the same as the middle panel. We observe a clear spiral arm pattern in the DM halo, strongly correlated with the stellar spiral arms.}
    \label{fig:density_cartesian}
\end{figure*}

\section{Simulations and methods}\label{sect:simulations_methods}

In this study, we analyse several simulations that present strong two-armed stellar spirals. The main analysis of this paper is based on a pure $N$-body model of a MW-like isolated galaxy, which we label \texttt{Iso-NBody}. This model is from our new suite of simulations run in the awarded project in the call from the Open Clouds for Research Environments (OCRE). The model is a slight modification of the L2 model in \cite{laporte2018sag}, with three components (cold exponential stellar disc, stellar \cite{hernquist1990halo} bulge, and a Hernquist DM halo) but with no perturber. The resulting galaxy model displays a strong bar and prominent two-armed stellar spirals, with a nearly constant pattern speed of $\Omega_\text{s}\,\sim\,20$\,\kmskpc, reaching a density contrast of $20\%$ at $R\,\sim\,6$\,kpc (see Sect.\,\ref{sect:res_iso_density}).

To complement the analysis, we also examine other simulations, including a test particle model (\texttt{TP}), a tidal $N$-body model with a Sagittarius-like perturber (\texttt{Sgr-NBody}), a high-resolution hydrodynamical simulation (\texttt{Hydro}), and a cosmological simulation from the Auriga project (\texttt{Au27}). These models provide additional insights and robustness into the formation and evolution of spiral arms under different physical conditions and simulation set-ups. Further details of the simulations are provided in Appendix\,\ref{app:simulations}.

We analysed the response of the DM halo to stellar spiral arms by focussing on the region $|Z| < 4\,\text{kpc}$ and $R < 30\,\text{kpc}$. The potential impact of the vertical cuts on our results is found to be small (see Appendix~\ref{app:z_cut}). Key measurables include: radially normalized density ($\delta\rho_r$), which is the fractional deviation of the local density (integrating over the vertical dimension) with respect to the average (over all azimuths) at the same radius (Eq.\,\ref{eq:delta_pr}), which clearly  emphasized the non-axisymmetric variations; mean radial and azimuthal velocities (\mVR, \mVphi); and vertical breathing motion (\VZb), which quantifies vertical velocity asymmetry between the upper and lower halves of the disc (Eq.\,\ref{eq:breathing}). \VZb{ }traces the compression and expansion caused by the quadrupoles in the potential (spiral arms and bar).
We used a Fourier decomposition to detect the spiral structure, with the $m=2$ mode tracking the phase ($\phi_2$) and amplitude ($\Sigma_2$) of the spirals. The spiral pattern speed ($\Omega_\text{s}$) was derived from a linear regression of the phase over a $\pm 0.1$\,Gyr window. Face-on images, smoothed with Gaussian kernels, were generated to aid interpretation. For simulations with perturbers, we applied a centro-symmetrization (rotate by $180^\circ$ and average with the original) to remove the dipole signatures coming from the response to the perturber, and highlight the quadrupole signature of the spiral arms. Further details about the methods used are provided in Appendix\,\ref{app:methods}.

\section{Dark spirals in a pure $N$-body simulation}\label{sect:results}

\subsection{Density}\label{sect:res_iso_density}

\begin{figure}
    \centering
    \includegraphics[width=.49\textwidth]{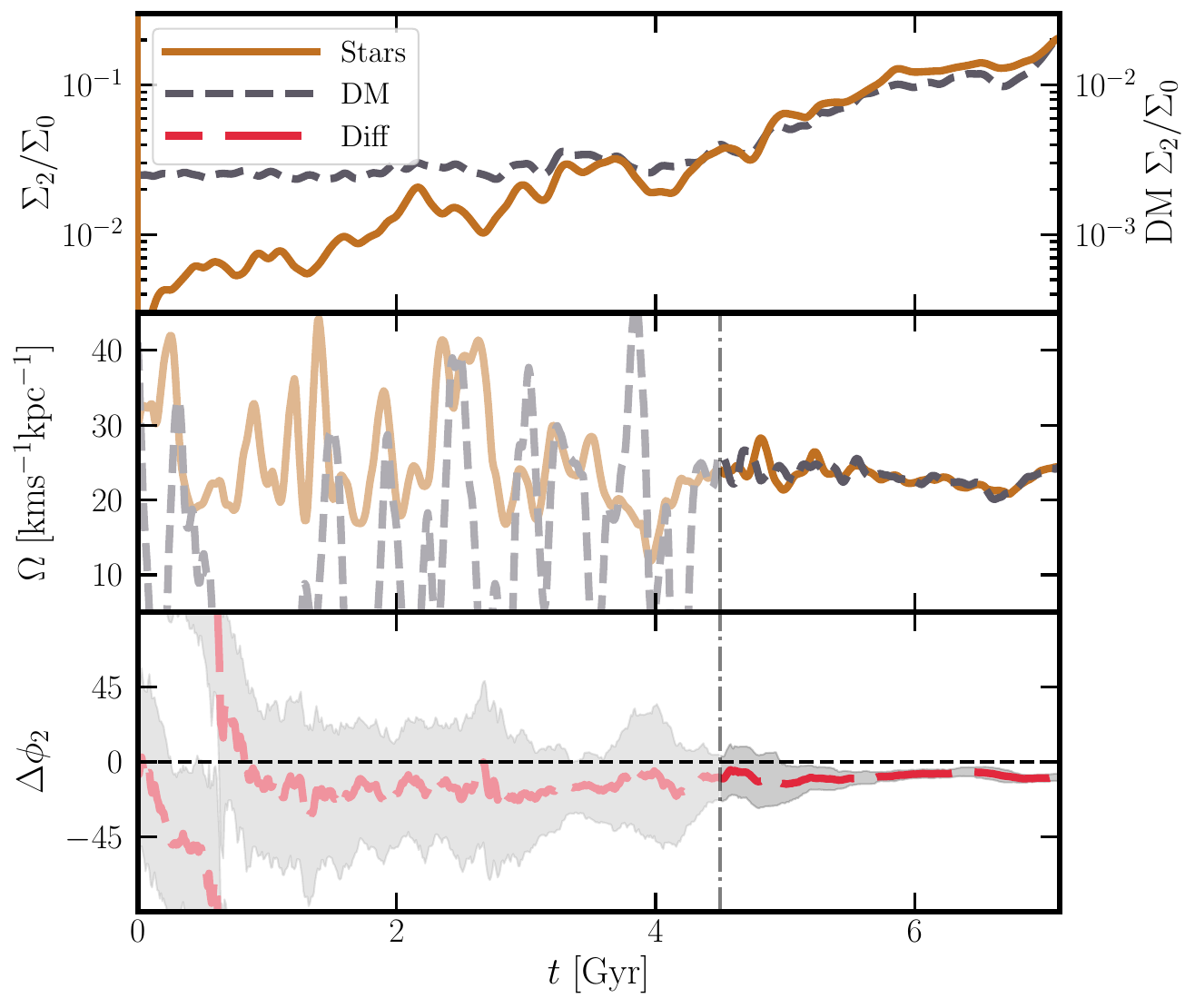}
    \caption{Temporal evolution of the \texttt{Iso-NBody} simulation in the range $R \,\in\,[5.5,6.5]$\,kpc. \textit{Top}: Relative amplitude of the Fourier mode ($\Sigma_2 / \Sigma_0$), i.e. strength of the spiral arms for the stellar (solid ochre) and DM (dashed grey) components. We see that both spirals evolve in a correlated way, although the relative amplitude of the DM spirals is one order of magnitude smaller (notice the auxiliary axis, which refers to the amplitude of the DM spiral arms). \textit{Middle}: Pattern speed ($\Omega_\text{s}$) of the mode $m=2$ structures in stars (solid ochre) and DM (dashed grey). \textit{Bottom}: Phase difference between the mode $m=2$ of the stellar and DM structures (long-dashed red line). Shaded regions shows the dispersion of the phase lag. Once the spirals are strong enough ($t \gtrsim 4.5$\,Gyr, vertical lines), we detect a constant pattern speed of $\Omega_\text{s}\,\sim\,20$\,\kmskpc{ }for the stellar and DM spirals, and a constant phase lag of $\Delta\phi_2\,\sim\,10^{\circ}$, showing a strong correlation between the two spiral arms.}
    \label{fig:density_time}
\end{figure}

We start by studying the \texttt{Iso-NBody} model. In the stellar density we clearly observe strong two-armed spirals reaching to $R\,\sim\,15$\,kpc with a peak $\delta\rho_r\,\sim\,0.2$ at $R\,\sim\,6$\,kpc (left and central panels of Fig.\,\ref{fig:density_cartesian}). Studying the temporal series (ochre lines in Fig.\,\ref{fig:density_time}) of the amplitude ($\Sigma_2 / \Sigma_0$) and phase ($\phi_2$), we observe that these spirals appear at $t\,\sim\,4.5$\,Gyr and grow with time with an approximately exponential profile to $\Sigma_2 / \Sigma_0\,\sim\,20\%$ at the end of the simulation. The pattern speed ($\Omega_\text{s}$) of the stellar spiral arms (central panel of Fig.\,\ref{fig:density_time}) is ill-defined before the appearance of the spiral, and stabilizes at $\Omega_s\,\sim\,20$\,\kmskpc{ } once the spiral emerges.

\begin{figure}
    \centering
    \includegraphics[width=.45\textwidth]{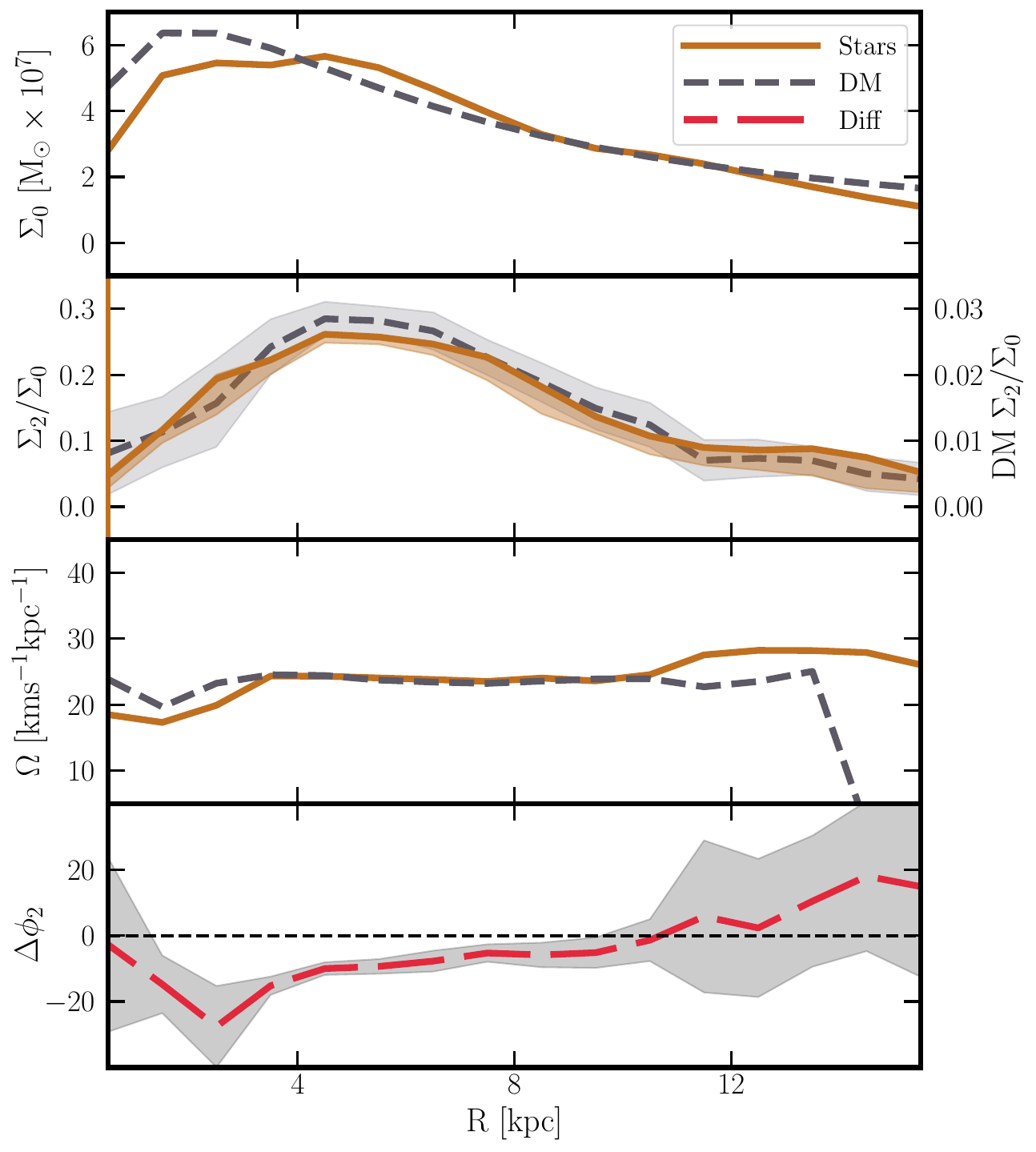}
    \caption{Radial profile of the \texttt{Iso-NBody} simulation at $t=7$\,Gyr. We show the Fourier amplitude and phase for the stellar (solid ochre) and DM (dashed grey) components. \textit{First}: Amplitude of the $0$th Fourier mode ($\Sigma_0$), in units of solar mass. We see that, close to the midplane, the surface density of stars and DM is comparable. \textit{Second}: Relative amplitude of the Fourier mode ($\Sigma_2 / \Sigma_0$) at all radii. The relative profile of the spirals is similar, with the relative amplitude of the DM spirals being one order of magnitude smaller than the stellar one (notice the auxiliary axis). \textit{Third}: Pattern speed ($\Omega_\text{s}$) of the mode $m=2$. \textit{Fourth}: Phase lag ($\Delta\phi_2$, long-dashed red line) between the DM and stellar components.}
    \label{fig:density_radii}
\end{figure}

Studying the radial profile of the stellar spiral arms at a given time ($t=7\,$Gyr, solid ochre lines in Fig.\,\ref{fig:density_radii}), we confirm that the strength of the spiral arms peaks at $R\,\sim\,6$\,kpc, and observe that their pattern speed is almost constant with the radius, going from the bar region up to the outer parts of the disc (solid ochre line in the third panel of Fig.\,\ref{fig:density_radii}).
For discussion about the pattern speed for different models, we refer to previous results \citep[e.g][]{grand2012corrotatingspirals,rocafabrega2013strongspiral,antoja2022l18}.

We now focus on the effect of the spirals on the DM halo (right panel in Fig.\,\ref{fig:density_cartesian}). We observe a DM spiral arm-shaped overdensity, nearly coincident with the position of the stellar spiral arms. The overdensity of the spiral arms in DM is $\delta\rho_r\,\sim\,0.02$, one order of magnitude smaller than that of the stellar ones.
In fact, the density contrast in DM spirals is roughly one order of magnitude lower than the density contrast of stellar spirals for all the models that we have studied (see Sect.\,\ref{sect:ubiquity}).

We see a strong correlation between the two spirals in terms of temporal evolution (dashed grey line in Fig.\,\ref{fig:density_time}, after $t\,\sim\,4.5$\,Gyr). The relative amplitude of the DM spirals (top panel) evolves linearly with the stellar one, and is about $10\%$ of the stellar relative amplitude at all times after $4.5$\,Gyr (as is indicated by the auxiliary vertical axis in the plot). The pattern speed of the spiral arms is also almost identical in the two components (stellar and DM) once the signal is strong enough ($t\,\sim\,4.5$\,Gyr). Finally, in the bottom panel of Fig.\,\ref{fig:density_time} we show the phase lag between the DM spiral arms and the stellar spiral (long-dashed red line). We observe that initially there is no correlation: the dispersion is $\sigma\,\sim\,45^\circ$, which is the expected dispersion for a uniform distribution. Once the stellar and DM spirals are formed, the DM spiral is consistently trailing the stellar spiral by $\sim\,10^\circ$.
This phase lag is a natural consequence of the delayed response of the DM halo to the evolving gravitational potential of the spiral arms, comparable to the mechanism at work in barred galaxies \citep[e.g.][]{weinberg1985slowbar,athanassoula2003bar}.

Finally, we focus on the radial profile of the DM spirals at $t=7$\,Gyr (Fig.\,\ref{fig:density_radii}). In the top panel, we notice that the local density ($\Sigma_0$) of stars and DM is very similar in this model. 
The relative amplitude, $\Sigma_2/\Sigma_0$, of the DM spiral arms (second panel) varies with radius but remains consistently $10\%$ of the stellar amplitude, as is shown by the secondary vertical axis on the right. We thus reveal a linear relationship between the stellar and DM amplitudes as a function of radius, matching the relationship observed in the temporal evolution (top panel of Fig.\,\ref{fig:density_time}).
This corresponds to adding about $10\%$ to the total mass of the stellar spiral arm, since $\Sigma_0$ of both the stellar and DM components are of the same order. Finally, we confirm the consistent pattern speed of the stellar and DM spiral arms across all radii (third panel in Fig.\,\ref{fig:density_radii}), and the lagging of the DM spiral with respect to the stellar spiral at all radii where the signal is strong (bottom panel in Fig.\,\ref{fig:density_radii}). We see a small trend of the phase lag with radius that is a consequence of the local properties of the stellar spiral arms and DM distribution function (DF), which one could potentially predict using perturbation theory.

\subsection{Velocity Space}

\begin{figure*}
    \centering
    \includegraphics[width=.9\textwidth]{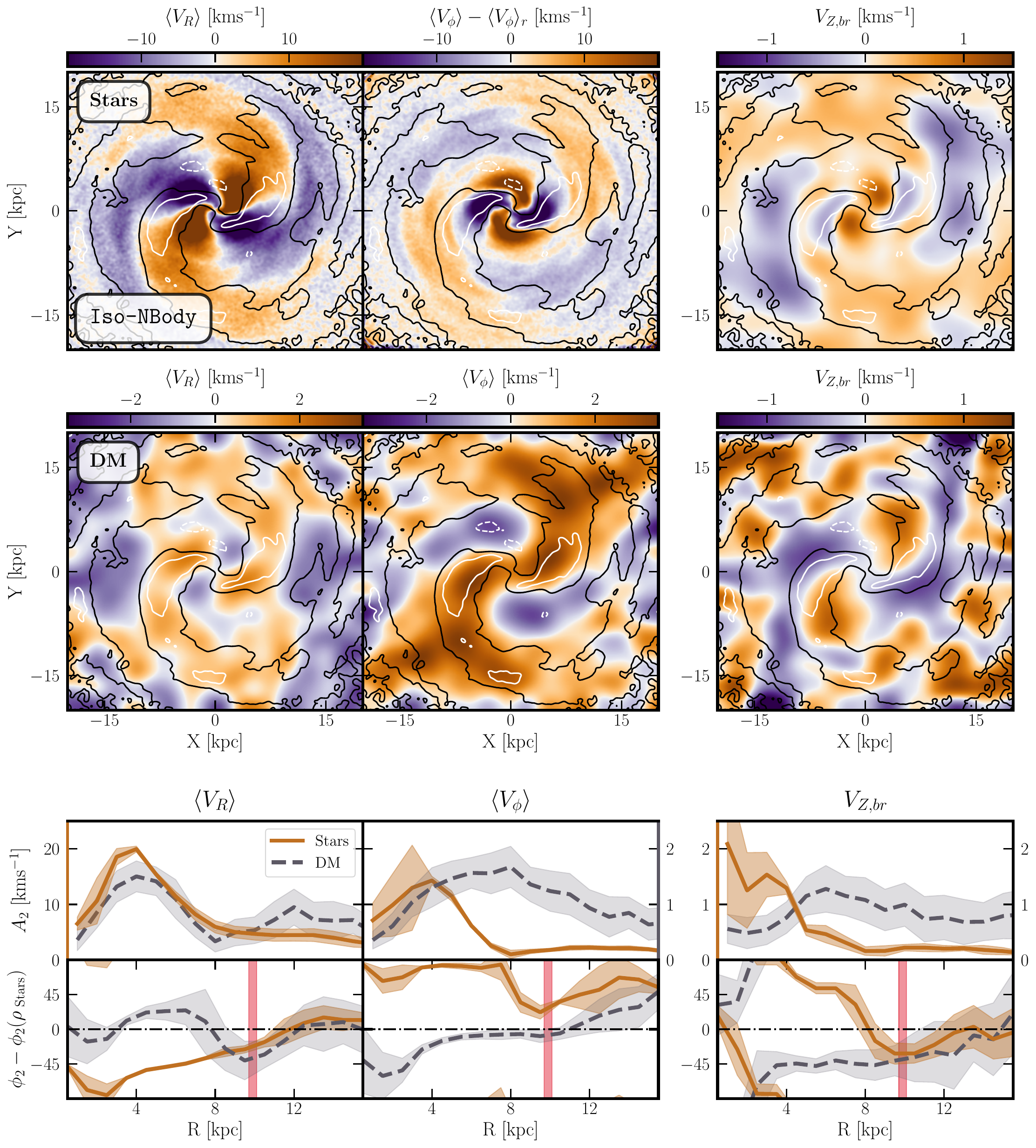}
    \caption{Kinematic analysis of the \texttt{Iso-NBody} simulation, at $t=7$\,Gyr. The contours represent the regions where $\delta\rho_r=0$ (in black) and $\delta\rho_r=\pm0.2$ (in white). We show \mVR{ }(left column), \mVphi{ }(middle column), and \VZb{ }(right column). \textit{First row}: Kinematic maps of the stellar component. \textit{Second row}: Kinematic maps of the DM component. \textit{Third row}: Radial profile of the kinematic amplitude of the Fourier mode ($A_2$) for the stellar (solid ochre) and DM (dashed grey) components. \textit{Fourth row}: Radial profile of the phase difference between each kinematic mode ($\phi_2$) and the stellar spiral in density ($\phi_2\,(\rho_{\text{Stars}})$). The co-rotation radii of the spirals ($R_{CR}\,\sim\,10$\,kpc) is shown as a vertical red strip.}
    \label{fig:velocity_cartesian}
\end{figure*}

In a dynamical system, the appearance of an overdensity in the configuration space must leave a trace in the kinematics. In this section we study \mVR, \mVphi, and \VZb{ }of the stellar and DM components in the \texttt{Iso-NBody} model.

We start by analysing the kinematic imprints in the stellar spirals. In the top row of Fig.\,\ref{fig:velocity_cartesian}, we show the kinematic maps of the stellar component, compared with the position of the bar and spiral arms (shown as contours in the figure). We observe three clear regimes: the bar, the inner disc, and the outer disc. In the third row of Fig.\,\ref{fig:velocity_cartesian}, we show the amplitude of the velocity quadrupoles at each radius (solid ochre line), and confirm that the in-plane velocity quadrupoles (\mVR and \mVphi) reach about $20\,\kms${ }in the inner part of the disc, and $5\,\kms${ }in the outer parts. As for the phase of the quadrupole (last row of Fig.\,\ref{fig:velocity_cartesian}), which we measure with respect to the phase of the stellar density structures, we observe a $90^\circ$ change in the phase of the quadrupole inside and outside the co-rotation\footnote{
The co-rotation radius was calculated based on the mean pattern speed within the range $R\,\in\,[4,8]$, using the rotation curve of this model derived from \textsc{Agama} \citep{vasiliev2019agama}.} of the spirals (vertical red line).
The kinematic signature of the spirals in our model is rather complex and depends on the radius and how the spirals connect with the bar at each time. However, since this is not the main focus of this article, we refer to the ongoing discussion on the topic \citep{siebert2012spiral,grand2015spiralkin,monari2016spiral_b,monari2016spiral_a,antoja2016spiralkinematics,antoja2022l18,eilers2020spiral_strength}.

\begin{figure*}
    \centering
    \includegraphics[width=1.\textwidth]{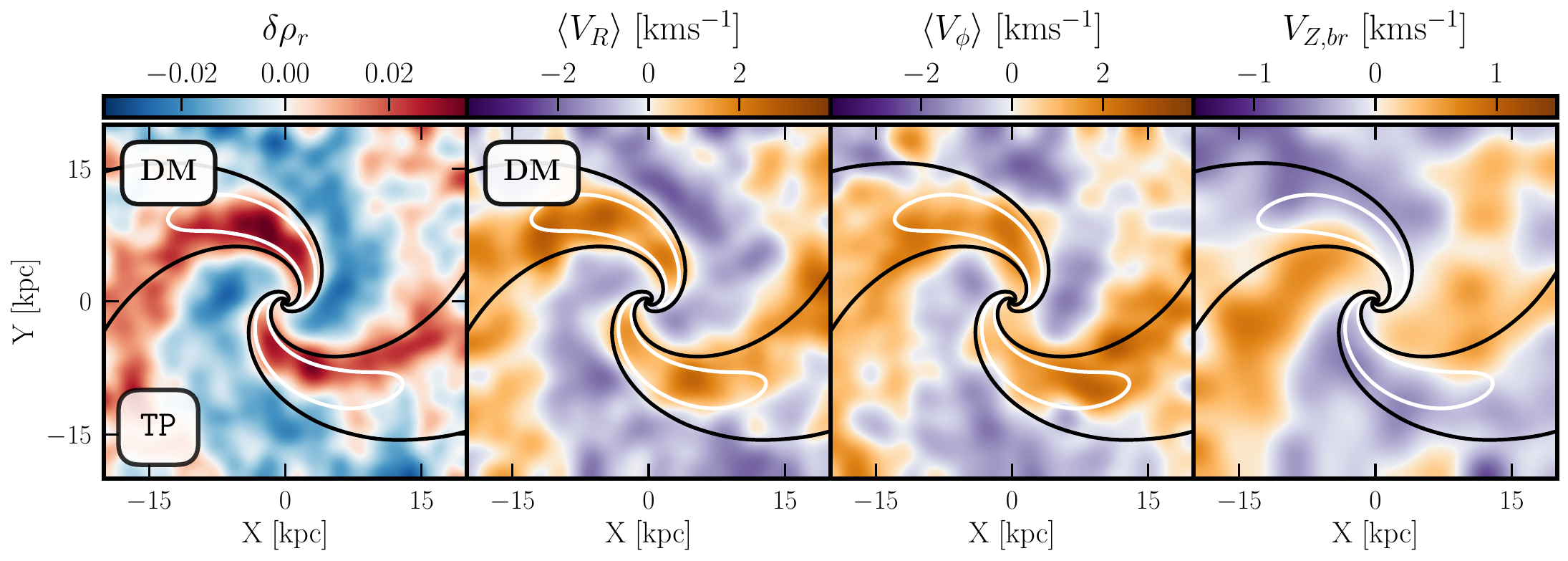}
    \caption{Response of the halo test particles to an analytical spiral potential \citep{cox2002spiral}, rotating at a constant pattern speed. From left to right, we show the radially normalized density ($\delta\rho_r$), the mean radial and azimuthal velocities (\mVR, \mVphi), and the vertical breathing motion (\VZb). In black and white, we show the $0$ and $0.5\rho_{max}$ density contours of the potential, respectively. The response of the halo is compatible with the results shown in Figs.\,\ref{fig:density_cartesian} and \ref{fig:velocity_cartesian}.}
    \label{fig:tp_spirals}
\end{figure*}

In the DM particles (second row in Fig.\,\ref{fig:velocity_cartesian}), we also observe quadrupoles in the kinematics. In \mVR{ }and \mVphi, we observe that along all the stellar spiral overdensity (solid contours) we have positive values, reaching amplitudes of $\sim\,2\,\kms$, much smaller than the stellar kinematic amplitudes. In the third row of Fig.\,\ref{fig:velocity_cartesian} (dashed grey lines), we show the amplitudes of the kinematic signatures. For \mVR, the amplitude of DM behaves qualitatively similar to the stellar one,  with a peak at $R\,\sim\,4$\,kpc, and lower values at outer radii. However, the radial profile of \mVphi{ }shows a different behaviour when comparing the stellar and the DM components, with the latter peaking at $R\,\sim\,8$\,kpc.
In the last row of Fig.\,\ref{fig:velocity_cartesian}, we show the phase of the velocity quadrupoles in the DM halo (dashed grey lines).
The distinct in-plane kinematic signatures between the stellar and DM spiral arms reflect their differing intrinsic dynamics: the stellar component, governed by rotationally dominated orbits, responds to the spiral potential through self-gravity and angular momentum exchange, while the DM halo, with its near-isotropic distribution, exhibits a kinematic response shaped primarily by the forced response rather than by self-gravity.

In the vertical response of the stellar component (\VZb, top right panel of Fig.\,\ref{fig:velocity_cartesian}), we also observe a clear quadrupole, but with a small amplitude of $\sim\,2\,\kms${ }in the inner parts of the disc, and $<1\,\kms${ }in the outer regions. It should be noted that \VZb{ }is tracing the compression and expansion caused by the spiral potential.
Outside co-rotation ($R>10$\,\kpc), the breathing motion is contracting (purple) in the leading part of the arm and expanding (orange) in the trailing. Inside co-rotation ($R<10$\,\kpc), where the stars orbit faster than the spiral arms and ‘overtake’ them, there is a change in the phasing of the vertical force that leads to a change in the sign of the quadrupole (phase shift of $90^\circ$ in the bottom panels). 
These findings are compatible with the predictions for the vertical kinematics induced by grand design spirals \citep{debattista2014breathing, faure2014breathing}.

The vertical response of the DM halo is shown in the third column of Fig.\,\ref{fig:velocity_cartesian}. In this case, the breathing motion contracts (purple) in the leading part of the arm and expands (orange) in the trailing part. This is similar to the stellar breathing motion only outside co-rotation. In a non-rotating halo, on average the particles encounter the spiral arm through the leading part at all radii, explaining the similarities between components of the vertical response outside co-rotation. Furthermore, in the third and fourth rows of the figure we observe that the amplitude ($\sim\,1\,\kms$) and phase of the DM mode are almost constant with radius.
While the differences in the planar motions of the stars and DM particles hint that we are observing dynamically different components, the similarities in the vertical breathing motions are not unexpected as they naturally result from the response of a passing overdensity corresponding to the spiral arms.

\section{Response to a fixed potential}\label{sect:results_tp}

As we discussed in the introduction, we expect that the dominating dynamical process creating the DM spiral arms is the `forced response' of the halo to the stellar spiral arms. To test this, we integrated halo orbits using a test particle set-up (\texttt{TP}; Appendix \ref{sect:sim_tp}), whereby halo orbits were integrated in an analytical galactic potential including spiral arms, isolating the forced response by removing self-gravity. By comparing the resulting density and kinematic signatures with those observed in the \texttt{Iso-NBody} model, we verified that the forced response remains identical even when self-gravity is absent.

In Fig.\,\ref{fig:tp_spirals}, we show the shape of the interaction signal in the four measurables we are studying. We see the clear correlation between DM and stars in density (first panel) and the slight lag of the DM spiral. For the kinematic modes, we see exactly the same patterns as for the \texttt{Iso-NBody} model. We observe clear quadrupoles in velocity, with positive \mVR{ }and \mVphi{ }on top of the spiral arms (second and third panels in Fig.\,\ref{fig:tp_spirals}), and amplitudes of $\sim\,2\,\kms$, in the same range as the ones observed in the \texttt{Iso-NBody} model. In the vertical breathing map, we also observe the change of sign in \VZb{ }on top of the spiral, with a breathing amplitude slightly smaller than the in-plane velocities. The signal approximately matches the one observed in \texttt{Iso-NBody}. Therefore, we conclude that the dominant dynamical process creating DM spiral arms in the \texttt{Iso-NBody} model is the forced response of the halo to the spiral arm potential.

We emphasize that no fine-tuning of the model parameters was required; we tested other combinations of spiral and MW-like model parameters and all produced a similar response.
Thus, we conclude that the precise simulation parameters have a second-order effect on the overall response of the halo, making our conclusions robust to these variations.

\subsection{Temporal response to the potential}

\begin{figure}
    \centering
    \includegraphics[width=0.5\textwidth]{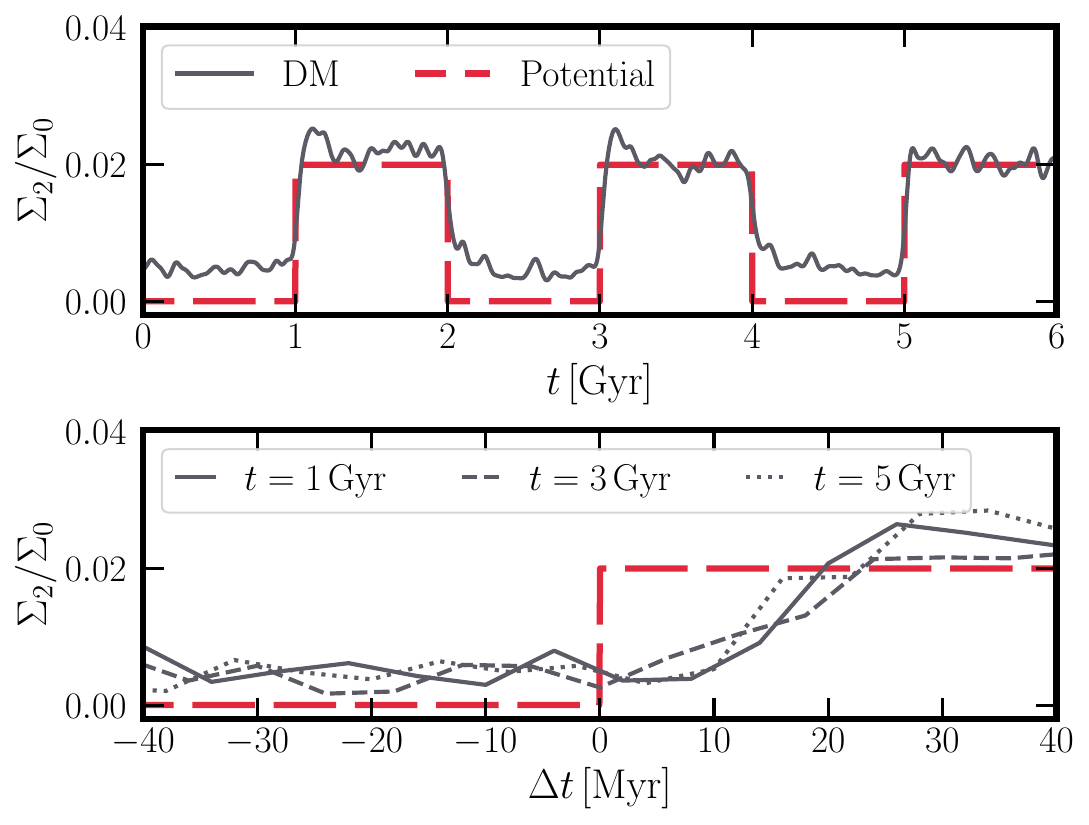}
    \caption{Strength ($\Sigma_2/\Sigma_0$, grey line) of the DM spiral arm at $R=8$\,kpc for the toy model.  In this set-up, the potential amplitude is switched on and off every $1$\,Gyr. The long-dashed red line represents the shape of the potential (its height is arbitrary, for illustration purposes). \textit{Top}: Whole evolution. \textit{Bottom}: Detail of the strength of the potential around the three increasing steps of the potential ($t=1,\,3,\,5$\,Gyr). We observe that the potential takes $\sim\,20$\,Myr to react to a change in the potential (impedance).}
    \label{fig:inout}
\end{figure}

The flexible \texttt{TP} set-up allows us to test the properties of the response that creates DM spiral arms, such as its temporal evolution. To do so, we conducted a different test in the same set-up as the \texttt{TP} simulation but in which, instead of a smooth rise in the spiral potential amplitude, we switched it on and off every $1$\,Gyr (Fig.\,\ref{fig:inout}). At each snapshot, we computed the strength ($\Sigma_2/\Sigma_0$) of the DM spiral arm (grey line in Fig.\,\ref{fig:inout}).
With this set-up, we can study the temporal delay (or the impedance, here referring to the resistance of the system to rapid change) of the response.
We observe that the response of the DM spirals to the sudden appearance or disappearance of a spiral potential is very fast ($\sim\,20$\,Myr, as is shown in the lower panel of Fig.\,\ref{fig:inout}).
This timescale closely matches the  local dynamical timescale given by the inverse epicyclic frequency ($1/\kappa\,\sim\,24$\,Myr) at $R=8$\,kpc in the \texttt{TP} model. This correspondence supports our interpretation that the response is governed by local orbital dynamics.

We also checked whether the presence of a long-lived spiral potential can align some halo orbits with the disc plane, increasing the presence of DM particles with disc-like orbits. These disc-aligned DM particles would be more sensitive to `traditional' spiral modes. To test this alignment, we extended the \texttt{TP} model to $t=6$\,Gyr. The fraction of DM particles with disc-like orbits increased by less than $1\%$, indicating that this alignment does not play a significant role.

\subsection{Response dependence on the orbital structure of the halo}\label{sect:orbital_distribution}

\begin{figure}
    \centering
    \includegraphics[width=.49\textwidth]{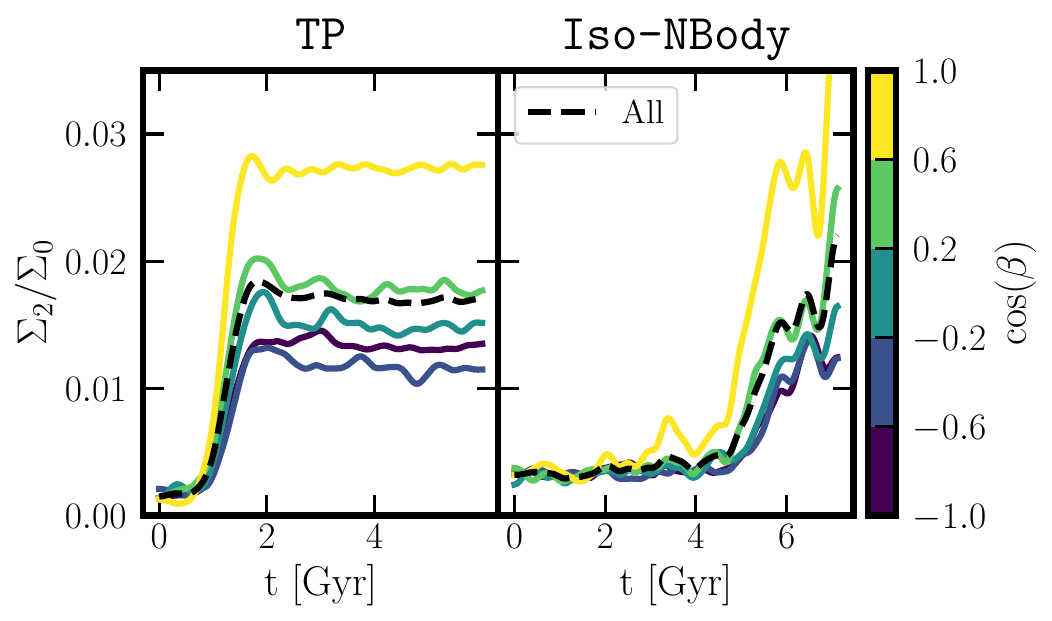}
    \caption{Strength ($\Sigma_2/\Sigma_0$) of the DM spiral arms at $R=6$\,kpc for the \texttt{TP} (left) and \texttt{Iso-NBody} (right) models. The DM halo particles are split according to $\beta$, the instantaneous angle between the $L_z$ of the orbit and the disc.}
    \label{fig:orbital}
\end{figure}

As we reviewed in the introduction, the interactions between the components of the Galaxy have a resonant nature \citep{tremaine1984slowbar}. Therefore, we expect the halo particles that are closer to co-rotation with the stellar spiral to be more affected by it.
In \citet{petersen2016shadowbar}, the efficiency of torque transfer is explained using a formal framework based on the torque expressions derived in \citet{weinberg1985slowbar}. Specifically, the LBK torque \citep{lynden-bell1972lbk} and non-linear torque \citep[Eqs.\,29 \& 41 in][]{weinberg1985slowbar} depend on rotation matrices, $R(\beta)$, where $\beta$ is the instantaneous angle between the angular momentum vectors ($L_z$) of the orbits and the disc plane. These expressions show that the torque transfer efficiency varies with $\beta$, with low-$\beta$ orbits being more sensitive to the torque than high-$\beta$ orbits.

In Fig.\,\ref{fig:orbital}, we test this difference in response by splitting the halo particles according to $\cos \beta$ in the \texttt{TP} and \texttt{Iso-NBody} simulations. At each snapshot, we evaluate the strength of the DM spiral arms for each sub-population. We clearly observe that the disc-like orbits in the halo ($\cos \beta \,\sim\, 1$) are about twice as sensitive to the stellar spirals as orbits that are counter-rotating or perpendicular to the disc.
It is worth noting, however, that even the DM particles that are counter-rotating with respect to the disc (dark lines in Fig.\,\ref{fig:orbital}) are influenced to some extent by the stellar spiral arms.

There is evidence that the DM halo of the MW is triaxial \citep[e.g.][]{han2022triaxialdm}, a common feature in MW-like galaxies, as has been shown by cosmological simulations \citep[e.g.][]{prada2019triaxialdm,dillamore2022triaxialdm,han2023cosmohalos}. This could influence the strength of the DM spiral arms.
To investigate the effect that rotating and/or flattened halos have on the response of the DM halo to the stellar spiral arms, we tentatively tried employing a novel \textsc{Agama} module. This tool enables the construction of self-consistent, rotating, and flattened DM halos by fitting a \texttt{DoublePowerLaw} DF to a target density profile\footnote{We checked a flattening $q\in(0.3,1.2)$, covering the expected range for cosmological halos seen both in simulations \citep[e.g.][]{chua2022oblateness} and observations \citep[e.g.][]{das2023oblateness}.}. We then performed the same orbit integration as in the \texttt{TP} model for each DF-potential pair generated. Flattening the halos unavoidably introduced changes in the DF to maintain the self-consistency, which made comparing the response of the flattened halos non-trivial.
On the other hand, prograde rotation clearly amplified the spiral arm response, while retrograde motion diminished it.

\section{Ubiquity of the dark spiral arms}\label{sect:ubiquity}

\begin{figure*}
    \centering
    \includegraphics[width=.96\textwidth]{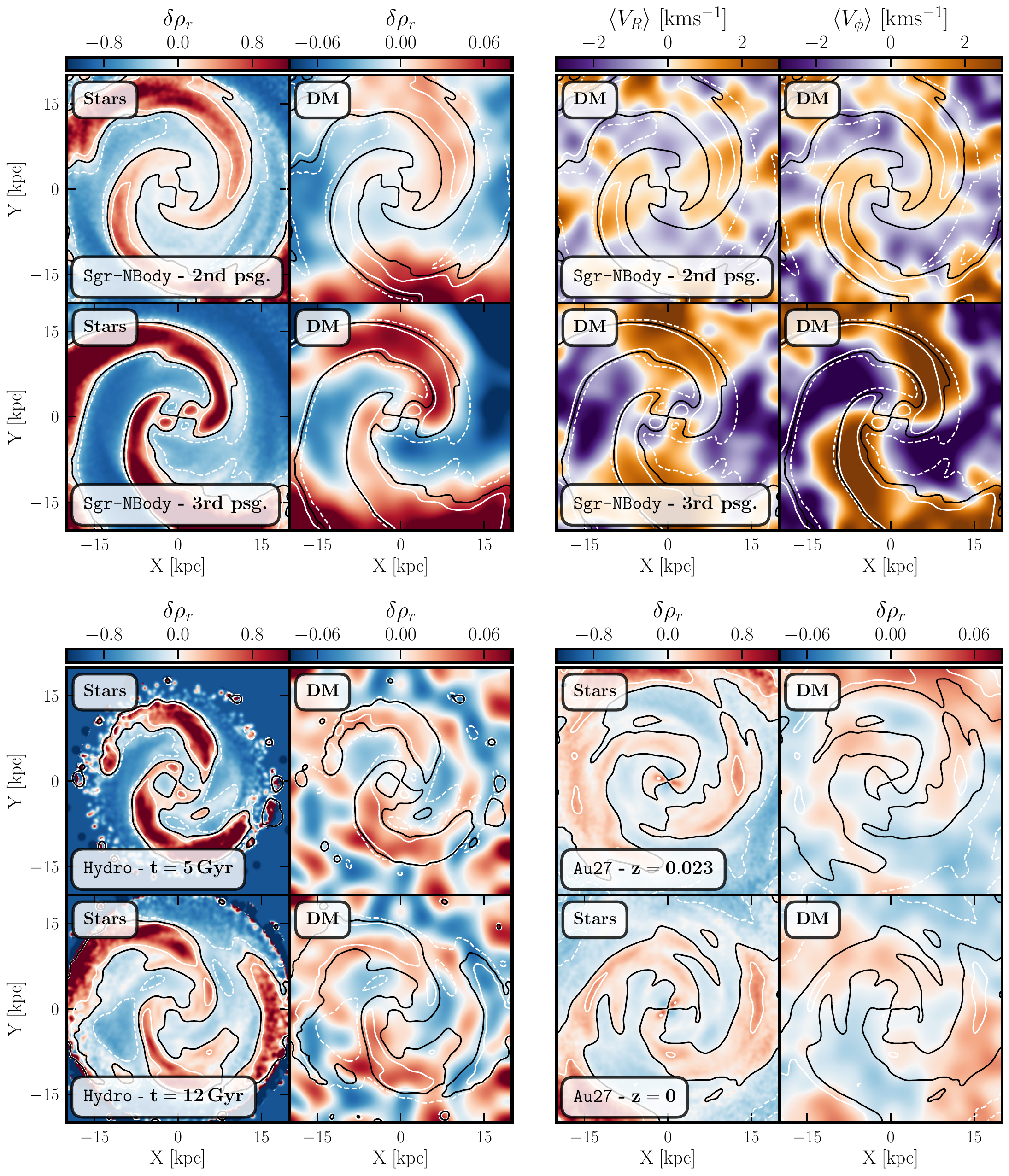}
    \caption{Dark matter spiral arms in other simulations. \texttt{Sgr-NBody} \textit{(upper two groups of panels)}: Stellar density (first column), DM density (second column), \mVR{ }(third column), and \mVphi{ }(fourth column) after the second and third passages of a Sgr-like perturber (first and second rows, respectively). \texttt{Hydro} \textit{(bottom left group)}: Stellar and DM density for two snapshots with strong spiral arms. \texttt{Au27} \textit{(bottom right group)}: Stellar and DM density for the last two snapshots in the simulation, $z=0$ and $z=0.023$. }
    \label{fig:selection}
\end{figure*}

To complement the analysis, we explored the presence of dark spiral arms in other simulations: a pure $N$-body with a Sgr-like perturber (\texttt{Sgr-NBody}), an isolated hydrodynamical (\texttt{Hydro}), and a cosmological (\texttt{Au27}), all of which are described in Appendix \ref{app:simulations}.
In Fig.\,\ref{fig:selection}, we show the results for these models. In the \texttt{Sgr-NBody} model, we have enough resolution to study the velocity maps. However, the number of particles and the lower strength of the stellar spiral in the models \texttt{Hydro} and \texttt{Au27} makes it very difficult to observe the kinematic signature of the DM spirals (Table\,\ref{tab:simulation_properties}).

The \texttt{Sgr-NBody} model reveals a clear two-armed spiral pattern after the second and third passages of the perturber. These stellar spiral arms have a tidal origin, and thus their dynamical characteristics, such as pattern speed and kinematic signatures, are different compared to the \texttt{Iso-NBody} \citep[e.g.][]{antoja2022l18}.
However, the observed response in the DM halo remains consistent with that seen in the \texttt{Iso-NBody} and \texttt{TP} models. We observe a clear DM spiral arm aligned with the stellar spiral arms and a similar kinematic signature to that of the \texttt{Iso-NBody} and \texttt{TP} models, with positive \mVR{ }and \mVphi{ }values on top of the spiral overdensity. In addition, we have studied the \texttt{Sgr-NBody} model with the same Fourier techniques as for the \texttt{Iso-NBody}, and obtained very similar results in terms of strength of the DM spirals and lag in phase. Thus, we conclude that the dynamical signatures of the forced response in the DM halo is independent of the dynamical origin of the stellar spiral arms.

To show the DM spiral arms in the Hydrodynamic simulation (\texttt{Hydro}), we selected an early snapshot ($t = 5$\,Gyr) and  one near the end of the simulation ($t=12$\,Gyr), which are the ones with a stronger two-armed spiral signature. In both cases, we observe signatures of the interaction between the dominant stellar spirals and the DM halo in the density space (bottom left groups of panels in Fig.\,\ref{fig:selection}). As in previous models, we observe the slight phase shift and the difference of about one order of magnitude in strength.

Finally, we show the $z=0.023$ and $z=0$ snapshots of an Auriga halo, $27$ (\texttt{Au27}, Fig.\,\ref{fig:selection}, bottom right quadrant). Both snapshots show a small bar with two grand design spirals, with a lower strength than the previously studied models ($\delta\rho_r\,\sim\,0.4$). Once again, we can identify the signatures of interaction in the density space quite clearly, shown by the spiral arm shaped overdensities in the halo, located on top of the stellar overdensity, with a slight lag in phase.

\section{Discussion}\label{sect:discussion}

\subsection{Formation mechanism}

Our results show that the DM halo responds to the perturbation caused by the stellar spiral arm potential. This is consistent with the predictions of the first-order perturbation theory \citep{mark1976}, which states that the response of the halo should be proportional to the mass of the spiral arms. The novelty is that with state-of-the-art simulations we can directly measure this interaction (Sects.\,\ref{sect:results}\,and\,\ref{sect:ubiquity}), and we can go a step further in studying its nature. We find a DM spiral amplitude of $\sim\,10\%$ of the stellar spiral amplitude, a consistent phase lag (trailing with respect to the stellar spirals by $\sim\,10^\circ$), and a characteristic kinematic signature in the halo.
We also show that the DM halo responds to the presence of a spiral density perturbation within $\sim\,20$\,Myr, a time compatible to the expected timescale, $1/\kappa$, and that DM particles with disc-like orbits are more likely to be affected by the spiral potential.

Throughout this work, we have shown evidence that the DM spiral arms are formed through the forced response of the DM halo to the stellar spiral arms.
First, the amplitude of the response of the halo is linear with respect to the amplitude of the stellar spiral arms in all the studied simulations. 
Second, with the TP (Sect.\,\ref{sect:results_tp}) we are able to reproduce the DM halo response to the spirals in both density and kinematic space. Since this model excludes self-gravity, it prevents the emergence of self-excited dynamical modes in the halo, meaning that we see a pure response to the perturbation. In addition, using simple test particle set-ups, we show the fast response of the halo and the absence of significant subsequent alignment of the orbits with the disc. Therefore, we conclude that the signal observed in our simulations is predominantly due to the forced response of the DM halo to the stellar spiral arms.

\subsection{Subdominant mechanisms}

While the forced response dominates, secondary mechanisms may also contribute to the DM spiral arms formation. Here, we examine two of them: self-excited halo modes and stellar bar imprints in the DM halo.

In an isolated, non-cosmological DM halo, the formation of DM spiral-arm structures as self-excited gravitational modes is unlikely. This is mainly due to two factors: the lack of coherent rotation within the halo and its high velocity dispersion. These prevent the alignment necessary to form spiral arms, making coherent patterns unsustainable.

For the second mechanism, we note that here we have used a barred spiral galaxy as our fiducial model (\texttt{Iso-NBody}). Therefore, there is a possibility that the presence of a bar is contributing to the appearance of DM spirals.
To test whether DM spiral arms could appear with `only' a bar, we modified the test particle set-up and changed the spiral arms potential \citep{cox2002spiral} for a \citet{portail2017bar} bar (Appendix\,\ref{app:tp_bar}). With this set-up, we are able to easily reproduce the shadow bar signature mentioned in the introduction.
The bar produces a spiral signature in density in the DM halo. However, the strength is one order of magnitude lower than the one observed when there is a spiral potential. In the kinematic space, there is a spiral signature in \mVphi, with a smaller amplitude and a much larger pitch angle than the spirals in \texttt{Iso-NBody}. In \mVR{ }and \VZb, there is no signature. 
In addition, we do not see evidence of DM particles following invariant manifolds \citep[e.g.][]{romerogomez2006manifold,voglis2006manifolds} in the DM halo.
Therefore, from this set-up we do not expect a significant direct contribution from the stellar bar to the appearance of the DM spiral arms of interest. 
Nonetheless, bars have been proposed to be able to induce spiral arms in the stellar disc \citep{dobbs2014spiralreview, sellwood2022review}, which would generate DM spiral arms, again through a forced response.

\subsection{Implications for the spiral arms morphology}

We have shown that, in MW-like halos, the relative amplitude of the DM spiral arms is consistently one order of magnitude smaller than the stellar spiral arms (Figs.\,\ref{fig:density_cartesian}, \ref{fig:density_time},\,\ref{fig:selection}). For instance, for a stellar spiral with a relative over-density of $20\%$ \citep[e.g][]{widmark2024dynamicalmass}, we predict a DM spiral arm with a relative over-density of $2\%$. This is translated into an increase of the total mass of the spiral arms of about $10\%$ due to the overdensity in the DM halo, very similar to the predictions of the contribution of the "shadow bar" \citep{petersen2016shadowbar}. In the planar kinematic space, we observe a similar relation: while the signature of a stellar spiral arm is $\approx 10\,\kms$, the measured kinematic signature of the dark spiral is $\approx 1\,\kms$ (Figs.\,\ref{fig:velocity_cartesian},\,\ref{fig:selection}).
Given these amplitudes, we expect the internal dynamics of the stellar spiral arms in the MW to dominate over the coupling with the DM halo.

However, these interaction effects are resonant, meaning that certain configurations amplify the DM spiral arm response.
In Sect.\,\ref{sect:orbital_distribution}, we show that DM particles on orbits aligned with the stellar disc are twice as sensitive to stellar spiral arms perturbations. We argue that this is consistent with analytical expectations: particles co-rotating with a spiral arm experience its gravitational influence over longer timescales. We also found that prograde rotating halos enhance the DM spiral response. 
While we have attempted to link the DM spiral response to the flattening of the halo, the interplay of the flattening with the other DM halo properties complicates the dynamical interpretation.
A future statistical analysis of DM spiral amplitudes in cosmological simulations -- correlating them with stellar spiral and halo properties -- would be an interesting follow-up project.

\section{Summary and conclusions}\label{sect:conclusions}

In this work, we present the first examination of the morphological and kinematic imprint of stellar spiral arms on the DM halo. We show that this interaction produces spiral-shaped overdensities in the DM halo. The appearance of a spiral overdensity in the halo is expected by a pure gravitational response. Our analysis confirms that the dominant mechanism forming these DM spiral arms is the forced response of the halo to the passage of the stellar spirals. Halo modes and bar forcing are not expected to form spirals with a density and kinematic signature consistent with the ones observed in the simulations.
Our main findings and conclusions are the following:

\begin{itemize}
    \item Whenever we observe stellar spiral arms (with $\Sigma_2/\Sigma_0\gtrsim5\%$), we see an increase in the total mass of the spiral arms of about $10\%$ due to the added mass from the DM halo response. The exact value is likely to depend on the properties of the spirals (e.g. pitch angle and pattern speed) and halo (e.g. rotation and anisotropy), and the region of the halo.
    \bigskip \item The phase of the kinematic signature of the spiral arm in the DM also correlates with the stellar spirals phase. However, the DM spiral arms show a distinctive quadrupole kinematic signature. On top of the spiral overdensity, we consistently observe positive \mVR{ }and \mVphi, and a change in the sign of the vertical breathing motion.
    \bigskip\item The amplitude of the planar kinematic signature in the DM halo is $\sim\,10\%$ that of the stellar component. However, the amplitude of the vertical breathing motion is similar in the stellar and DM components.
    \bigskip\item The interaction between the stellar spiral arms and the DM halo is a resonant effect. The DM particles rotating with disc-like orbits are more susceptible to the torque produced by the spiral potential (Fig.\,\ref{fig:orbital}). However, all the orbital families of DM particles are somehow affected by the interaction and produce spiral arm-shaped overdensities.
    \bigskip\item We show the presence of DM spiral arms in a broad range of different state-of-the-art simulations, indicating that DM spiral arms are a common feature in simulated MW-like galaxies.
\end{itemize}

With this work, we contribute to the understanding of a surprisingly overlooked dynamical phenomenon: the interaction between spiral arms and the DM halo. The expected signal in density and kinematics is faint and convolved with the stellar spirals. However, this interaction occurs close to the Sun, offering a potential probe for exploring the nature of DM.
If we were able to measure both the dynamical and baryonic mass distribution across a large region the MW disc, the difference between those would reveal the spiral substructure of the DM halo, rather than the expected first-order axisymmetric (or triaxial) distribution.
Infrared data modelled by \citet{drimmel2001spiral} suggest a two-armed logarithmic spiral with a relative strength of $10\%$, yet different samples and techniques produce results that can vary from $10$ to $20\%$ \citep[e.g.][]{eilers2020spiral_strength,khanna2024spiral}. As for the dynamical mass, \citet{widmark2024dynamicalmass}, using the vertical Jeans equation and \emph{Gaia} DR3 data, measured the distributions in the disc and inferred a local relative over-density of roughly $20\%$. The uncertainty in the current measurements of both the baryonic and dynamical mass makes the detection of DM spiral arms in the MW challenging but this could improve in the future. Once we reach that point, detailed predictions of the dark contributions will allow us to use the spiral arms as new tools to explore the nature of DM.

Hopefully, though, continuous improvements in the precision of the kinematic measurements of the stars and in the completeness of the samples used, aided by our ever-improving understanding of the baryonic dynamics, will allow us to confirm the predictions of this work. Once we reach that point, we shall be able to use the spiral arms as new tools to explore the nature of DM.

\begin{acknowledgements}
We thank the anonymous referee for her/his helpful comments. The authors thank Eugene Vasiliev for his support on the usage of \textsc{Agama}. We also thank Jason A. S. Hunt, the EXP collaboration, especially the Bars \& Spirals working group, and the Gaia UB team for the valuable discussions that contributed to this paper.
The OCRE simulations were run in virtual machines in the cloud provided by the OCRE awarded project Galactic Research in Cloud Services (Galactic RainCloudS). OCRE receives funding from the European Union’s Horizon 2020 research and innovation programme under grant agreement no. 824079.
The \texttt{M1\_c\_b} (\texttt{Hydro}) simulation was run at the DiRAC Shared Memory Processing system at the University of Cambridge, operated by the COSMOS Project at the Department of Applied Mathematics and Theoretical Physics on behalf of the STFC DiRAC HPC Facility (www.dirac.ac.uk). This equipment was funded by BIS National E-infrastructure capital grant ST/J005673/1, STFC capital grant ST/H008586/1 and STFC DiRAC Operations grant vST/K00333X/1. DiRAC is part of the National E-Infrastructure.
We acknowledge the grants PID2021-125451NA-I00, PID2021-122842OB-C21 and CNS2022-135232 funded by MICIU/AEI/10.13039/501100011033 and by ``ERDF A way of making Europe’’, by the ``European Union'' and by the ``European Union Next Generation EU/PRTR'', and the Institute of Cosmos Sciences University of Barcelona (ICCUB, Unidad de Excelencia ’Mar{\'\i}a de Maeztu’) through grant CEX2019-000918-M.
MB acknowledges funding from the University of Barcelona’s official doctoral program for the development of a R+D+i project under the PREDOCS-UB grant.
TA acknowledges the grant RYC2018-025968-I funded by MCIN/AEI/10.13039/501100011033 and by ``ESF Investing in your future''. 
JA is supported by the National Natural Science Foundation of China under grant No. 12233001, by the National Key R\&D Program of China under grant No. 2024YFA1611602, by a Shanghai Natural Science Research Grant (24ZR1491200), by the ``111'' project of the Ministry of Education under grant No. B20019, and by the China Manned Space Project with No. CMS-CSST-2021-B03. JA thanks the sponsorship from Yangyang Development Fund. J.A. \& C.L. acknowledge funding from the European Research Council (ERC) under the European Union’s Horizon 2020 research and innovation programme (grant agreement No. 852839).
RG acknowledges financial support from an STFC Ernest Rutherford Fellowship (ST/W003643/1). 
OJA acknowledges funding from ``Swedish National Space Agency 2023-00154 David Hobbs The GaiaNIR Mission'' and ``Swedish National Space Agency 2023-00137 David Hobbs The Extended Gaia Mission''.  
MSP is supported by a UKRI Stephen Hawking Fellowship. 
SRF acknowledges support from a Spanish postdoctoral fellowship, under grant number 2017-T2/TIC-5592.
\end{acknowledgements}

\bibliographystyle{aa}
\bibliography{MyBib}

%\newpage

\appendix

\section{Simulations}\label{app:simulations}

In this section we describe the different simulation set-ups and the main characteristics of each model, summarized in Table\,\ref{tab:simulation_properties}.

\begin{table}[]
\renewcommand{\arraystretch}{1.3}
\caption{Properties of the simulations. }
\label{tab:simulation_properties}
\begin{tabular}{ccccc}
\hline \hline
Label                & $N_{\text{disc}}$   & $M_{\text{disc}}$ [M$_\odot$] & $N_{\text{halo}}$ & $M_{\text{halo}}$ [M$_\odot$] \\
    & $\times10^6$   & $\times10^{10}$ & $\times10^7$ & $\times10^{12}$\\ \hline
$\texttt{Iso-NBody}$ & $10$ & $6$         & $4$          & $1$                    \\
$\texttt{TP}$        & -            & $4.56$         & $17$         & $1$                    \\
$\texttt{Sgr-NBody}$ & $5$ & $6$         & $4$          & $1$                     \\
$\texttt{Hydro}$     & $11$ & $6.47$        &   $0.5$        & $1$                    \\
$\texttt{Au27}$      & $15$ & $9$         & $1.9$          & $1$                   \\ \hline
\end{tabular}
\tablefoot{The columns give the label of the model, the number of particles in the disc ($N_{\text{disc}}$), the mass of the disc ($M_{\text{disc}}$), the number of particles in the halo, ($N_{\text{halo}}$), and the mass of the halo ($M_{\text{halo}}$).}
\end{table}

\subsection{Isolated disc - \texttt{Iso-NBody}}\label{sect:sim_iso-nbody}

The main analysis of this paper is based on a pure $N$-body model of a MW-like isolated galaxy. For the DM halo, a \citet{hernquist1990halo} profile with a scale length of $R_h = 52$\,kpc and a mass of $M_h = 10^{12}$\,M$_\odot$ is used. The initial stellar disc is exponential, with scale length $R_d = 3.5$\,kpc, scale height $h_d = 0.53$\,kpc, and total mass $M_d = 6\times10^{10}$\,M$_\odot$. Finally, a stellar bulge with $M_b = 10^{10}$\,M$_\odot$ and a Hernquist profile with a scale length of $0.7$\,kpc is included. The ratio $\sigma_R / \sigma_z = 1.5$ is fixed, so that the Toomre $Q$ parameter is $\sim$ 1 in the disc (i.e. $R\in[3,10]$\,kpc).
The initial conditions were generated using GalIC \citep{yurin2014galIC}, and evolved using the AREPO code \citep{weinberger2020arepo} for $7.12$\,Gyr. 

\subsection{Test particle (\texttt{TP})}\label{sect:sim_tp}

To compare the $N$-body response with a controlled set-up, we used a suite of test particle simulations.
We constructed a self-consistent model of a galaxy in equilibrium using \textsc{Agama} \citep{vasiliev2019agama}. We used a \citet{mcmillan2017potential} potential, and modelled the DM halo DFs as \texttt{DoublePowerLaw} \citep[][implemented in \textsc{Agama}]{binney2014df, posti2015doublepowlaw}\footnote{Extracted from the \texttt{example\_self\_consistent\_model.py} script of \textsc{Agama}.}. 
Finally, we iterate the model using the self-consistent-modeling \citep{binney2014scm,piffl2015scm} module of \textsc{Agama} to obtain a self-consistent DF of the halo.

Once the potential and DF are set up, we include a \citet{cox2002spiral} spiral arm model (with a null total mass),
\begin{equation}
    \Phi_{\rm{s}}(R,\phi,z)\!=\!-4\pi G \Sigma_0(t)\mathrm{e}^{-R/R_{\rm{s}}}\!\!\sum_{n=1}^3\!\frac{C_n}{K_n\,D_n}\!
    \cos n\gamma \big[\!\cosh\!\big(\tfrac{K_nz}{\beta_n}\big)\!\big]^{-\beta_n}\!\!,
    \label{eq:spiralarm}
\end{equation}
where $\Sigma_0$ is the central surface density, $R_{\rm{s}}$ is the scale radius, and $C_n$ are the amplitudes of the harmonic terms. The functional parameters are:
\begin{equation}
    \begin{aligned}
    &K_n\,=\,\frac{nN}{R\sin{\alpha}},\\
    &\beta_n\,=\,K_n\,h_{\rm{s}}(1+0.4K_n h_{\rm{s}}),\\
    &\gamma\,=\,N\bigg[\phi\,-\,\frac{\ln{(R/R_{\rm{s}})}}{\tan{\alpha}}\,-\,\Omega_{\rm{p}} t\,-\,\phi_0 \bigg],\\
    &D_n\,=\, \frac{1}{1+0.3K_nh_{\rm{s}}}\,+\,K_nh_{\rm{s}},
    \end{aligned}
    \label{eq:spiral_params}
\end{equation}
where $N$ is the number of arms, $h_{\rm{s}}$ is the scale height, $\alpha$ is the pitch angle, and $\phi_0$ is the initial phase. 

Following \citet{dehnen2000olr}, the spiral amplitude is initialized at $\Sigma_0(t) = 0$ for $t=0$, then increases smoothly between $t = 0.5$\,Gyr and $t=2$\,Gyr as
\begin{equation}\label{eqn:smooth_time}
    \Sigma_0(t) = \Sigma_{0,f}\biggl[\frac{3}{16}\xi^5-\frac{5}{8}\xi^3+\frac{15}{16}\xi+\frac{1}{2}\biggr], \quad \xi=2\frac{t}{1.5}-1,
\end{equation}
\noindent and it remains at a fixed $\Sigma_{0,f}$ for $t>2$\,Gyr until the end of the simulation at $t = 6$\,Gyr.

In this run, we select parameters that visually approximated the shape and amplitude of the stellar spirals in the \texttt{Iso-NBody} model: $N=2$, $\alpha=\pi/2$, $R_{\rm{s}} = 4$\,kpc, $C_n=[8/(3\pi)$, $1/2$, $8/(15\pi)]$, $h_{\rm{s}}=1$\,kpc, $\Sigma_{0,f}=2.7\times 10^8$\,M$_\odot$\,kpc$^{-2}$, $\phi_0=0$, and $\Omega_\text{s}=20$\,\kmskpc. We integrate $N_{\rm{halo}} = 1.7\times 10^8$ particles for $6$\,Gyr, sampling every $30$\,Myr.

For completeness, we rerun the same model using a \citet{portail2017bar} bar instead of spiral arm potential. Details of the model and results are presented in Appendix\,\ref{app:tp_bar}. In brief, we do see signatures of interaction between the DM halo and the bar, but the signature in configuration and kinematic space is very different to the observed for the spiral arms.

\subsection{Tidal pure $N$-body (\texttt{Sgr-NBody})}\label{sect:sim_sgr-nbody}

To test the interaction between the spiral arms and the DM halo in a perturbed galaxy, we also studied a pure $N$-body model from the same OCRE suite as \texttt{Iso-NBody} with a Sagittarius (Sgr)-like dwarf spheroidal galaxy perturber. Minor mergers like these can excite spiral arms and bar formation \citep[e.g.][]{toomre1972spiral,barnes1992spiral,dobbs2014spiralreview,sellwood2022review,jimenezarranz2024lmc}, and are thought to be causing some of the substructure observed in the \emph{Gaia} data \citep[e.g.][]{antoja2018phasespiral,binney2018phasespiral,laporte2019gaiaphase}.

The MW model is very similar to \texttt{Iso-NBody}, with a Sgr-like satellite and a slightly hotter disc. The orbit of the perturber and the MW properties of the simulation are almost identical to the L2 model in \citet{laporte2018sag}, while the Sgr-like perturber is modified to have a less concentrated DM halo, resulting in an increased mass loss history, more compatible to current observations (Laporte et al. in prep.).

From this simulation, we select two snapshots, at $t=4.94$, and $t=6.06$\,Gyr, located slightly after the second and third pericentres of the perturber. These are the moments where the simulation contains stronger spirals (see \citealt{grionfilho2021l18} and \citealt{antoja2022l18} for a detailed description of the spirals in L2).

\subsection{Isolated hydrodynamical (\texttt{Hydro})}\label{sect:sim_hydro}

We complement the set-up of pure $N$-body models with a self-consistent, high-resolution simulation with star formation. The model we use is a hydrodynamic MW-like model, initially introduced in \citet{fiteni2021hydro}, as model \texttt{M1\_c\_b}. This model develops through the cooling of a hot gas corona inside a DM halo. From this model, we select two snapshots with strong spiral arms ($t=5$, and $t=12$\,Gyr).

\citet{roskar2012hydro_ics} investigate the general evolution of the spiral structure using a lower-resolution version of the same model. They identify different coexisting pattern speeds, some of which are comparable to those obtained in the model \texttt{Iso-NBody} ($\Omega_\text{s} \,\sim\, 20$\,\kmskpc). This is confirmed in \citet{debattista2024hydroanalysis}, which uses the high resolution model \texttt{M1\_c\_b}. Finally, \citet{ghosh2022hydrospiral} used one of the snapshots we use ($t=12$\,Gyr) to study the breathing motions induced by its prominent spiral arms.

\subsection{Cosmological (\texttt{Au27})}\label{sect:sim_auriga}

Finally, we complement our analysis with a simulation from the Auriga project, in order to test the presence of DM spiral arms in cosmological state-of-art simulations. The Auriga project \citep{grand2017auriga} consists of 30 zoom-in cosmological magnetohydrodynamical high-resolution simulations of MW-mass dark halos, that have been proved to form realistic spirals \citep{grand2021realisticspirals}.
Using the latest public data release of the augmented Auriga Project \citep{grand2024aurigapublic}, we select the \texttt{Original/3} version of the \texttt{Au27} halo.
From the \texttt{Au27} halo, we select the two last snapshots ($z = 0.023$, and $z = 0$), which present a MW-like galaxy model with strong spiral arms.

\section{Methods}\label{app:methods}

In this section we describe the methodological details of each step.

\subsection{DM halo selection}\label{app:z_cut}

We are interested in the response of the DM halo to the presence of stellar spiral arms. Throughout this paper, when referring to the DM halo, we will be applying the following cuts:
\begin{equation}
    |Z|<4\,\text{kpc},\quad R<30\,\text{kpc}.
\end{equation}

These cuts are applied to increase the signal of the faint interaction we are studying. 
We tested modifications of the vertical cut over the range $|Z| \in (0.2,\,20)$\,kpc. The overall morphology of the DM spiral arms remains very similar throughout these tests. Notably, when selecting DM particles close to the disk (i.e., $|Z| \lesssim 1$\,kpc), the DM spiral signal can increase by up to $40\%$. However, this increase does not alter our conclusions, as the amplitude remains approximately one order of magnitude lower than that of the stellar spiral. Additionally, a narrower vertical range reduces the number of particles, thereby degrading the signal quality. For these reasons, we adopted a compromise by maintaining the cut at 
$|Z| < 4$\,kpc.

\subsection{Measurables}

\paragraph{Radially normalized density ($\delta\rho_r$):}

To highlight the non-axisymmetric features, we normalize the particle count in each bin by the mean density computed at that radius, where the average is taken over all azimuthal angles (integrating over the vertical dimension). This is defined as:
\begin{equation}\label{eq:delta_pr}
    \delta\rho_r (x,y) = N(x,y) / \bar{N}(R) - 1
\end{equation}
with $R = \sqrt{x^2 + y^2}$ and $\bar{N}(R)$ representing the mean particle count at radius $R$ calculated by averaging over all azimuthal angles. This normalization was used only in the density maps. The same result is achieved with the $0$-th order mode of the Fourier analysis.

\paragraph{Mean radial velocity \mVR\, and mean azimuthal velocity \mVphi:}

To study the overall dynamical properties of the samples, we focus on the mean velocities.
In the case of the \mVphi, in the planar maps we subtract the mean $V_\phi$ at each radius (\mVphi$_r$) for visualization and direct comparison with the other quantities. This subtraction is not done in the Fourier transform, since, again, $V_c$ is captured by the $0$-th order of the Fourier analysis.

\paragraph{Vertical breathing motion (\VZb):}

To study the vertical response to the spiral arms, we compute the breathing motion of the disc, which is the vertical asymmetry of the vertical velocity:
\begin{equation}\label{eq:breathing}
    V_{\rm{Z,br}}(x, y) = \frac{1}{2}\big(\langle V_Z(x,y,z_+) \rangle - \langle V_Z(x,y,z_-) \rangle\big)
\end{equation}
where $z_+ = z \in [0,4]$\,kpc, and $z_- = z \in [-4,0]$\,kpc \citep{debattista2014breathing,widrow2014breathing}. The vertical breathing motion will be positive in the regions where the disc is expanding vertically, and negative where the disc is contracting.

\subsection{Fourier decomposition}

We also use a Fourier analysis at each radius and snapshot as our quantitative measure to detect spirals. We split each snapshot in radial bins of $1$\,kpc with $100$ bins in azimuth. For each bin in radius, we compute the 1D Fourier transform of the azimuthal profile of $\rho$, \mVR, \mVphi, and \VZb. Since we are studying systems with strong, two-armed spirals, the $m=2$ mode is the dominant one and we use it to track the phase ($\phi_2$) and amplitude ($\Sigma_2$) of the spiral. When studying the density, we refer to the relative amplitude of the $m=2$ mode ($\Sigma_2 / \Sigma_0$). For the kinematic modes, we refer directly to the amplitude of the $2$nd mode ($A_2$). To compute the errors of the Fourier amplitude and phase, we use a window of $\pm 0.1$\,Gyr ($20$ snapshots) around the given time and compute the standard deviation of the value in the window.

We compute the pattern speed of the stellar and DM spiral arms in the simulation \texttt{Iso-NBody}. To compute the pattern speed ($\Omega_\text{s}$) of the $m=2$ mode at a given time, we select the $\phi_2$ values in a $\pm 0.1$\,Gyr window. We unwrap the phase values to avoid the $2\pi$ discontinuity and compute the linear regression of $t-\phi_2$ in the window. The slope of the regression is the pattern speed at the center of the window, once it is transformed to have \kmskpc{ }units. This method can be unstable if more than one spiral is present. However, in the simulation where we compute the pattern speed (\texttt{Iso-NBody}) there is only one pair of strong spiral arms and this method provides robust and accurate results.

\subsection{Face-on images}

We include face-on representations of the studied quantities to help the physical interpretation of the Fourier results. We smooth the images using a Gaussian kernel of width $0.3$\,kpc for the stellar component and $0.6$\,kpc for the DM component.

In the simulations with perturbers (\texttt{Sgr-NBody} and \texttt{Au27}), the motions in the DM halo (\mVR, \mVphi, \VZb) are dominated by the presence of a dipole in velocity \citep{weinberg1994m1mode, heggie2020m1mode,johnson2023bfe}. Since we focus on the study of the features with quadrupole signatures (two armed spiral arms), in Sect.\,\ref{sect:ubiquity} we show the centro-symmetrized component of the velocity maps of \texttt{Sgr-NBody} (average each pixel with its antipodal). This is
\begin{equation}
    \widetilde{A}_{i,j} = \frac{1}{2} \big(A_{i,j} + A_{n-i, n-j}\big) \quad \text{ for all } i, j \in {0,...,n-1}
\end{equation}
where $A$ is the 2D ($n\times n$) grid of the studied quantity in Cartesian coordinates, $i,j$ are the indices in the grid, and $\widetilde{A}$ its centro-symmetrized component. In practice, we are blindly removing all the odd modes of the image, highlighting the features we are interested in.

\section{Test particle simulation with bar}\label{app:tp_bar}

\begin{figure}
    \centering
    \includegraphics[width=.98\textwidth]{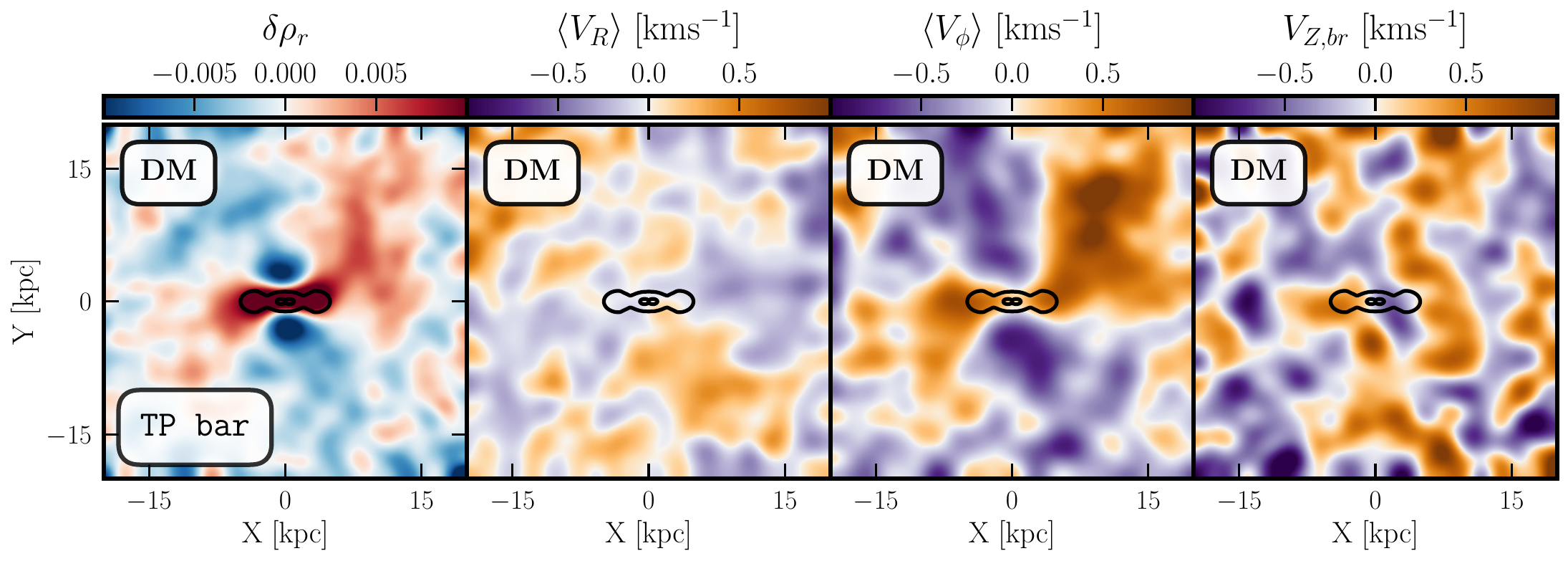}
    \caption{Response of the halo test particles to an analytical bar potential \citep{portail2017bar}, rotating at a constant pattern speed. In black, we show the $0.1$ and $0.5\rho_{max}$ density contours of the bar potential.}
    \label{fig:tp_bar}
\end{figure}

In this section we present the results of a test particle simulation with a purely barred potential to test if we can create DM spiral arms in this set-up.
We start with the exact same set-up as the one described in Sect.\,\ref{sect:sim_tp}, constructing a self-consistent model of a galaxy in equilibrium using \textsc{Agama}. In that set-up, we then include a \citet{portail2017bar} bar, rotating at $\Omega_{\rm{b}} = 20$\,\kmskpc, and growing with the same law as Eq.\,\ref{eqn:smooth_time}.

The results are shown in Fig.\,\ref{fig:tp_bar}. In $\delta\rho_r$ (first panel), we observe the clear presence of the shadow bar, and faint hints of the presence of spiral arms in the outer parts of the disc, with a strength below $0.1\%$. In \mVphi { }(third panel in Fig.\,\ref{fig:tp_bar}), there is a clear spiral kinematic signature that could resemble the one observed in the pure $N$-body models. However, the amplitude of this kinematic signature is maximum in the inner part, and significantly smaller than the observed in the other simulations. Finally, we observe hints of a quadrupole structure in \mVR, very different to the signature observed in the other models, and we do not observe any signal in \VZb.

In conclusion, we are able to produce substructure in the DM halo when having only a bar, as expected from the results in \citet{dillamore2023barhalo1,dillamore2024barhalo2}. In addition, this substructure is shaped as spiral arms. However, the signature in density and kinematics does not resemble the spiral signature we are studying.

\end{document}